\documentclass[journal]{IEEEtran}
\usepackage{cite}

\usepackage{SIunits}
\usepackage{float}

\newcommand\Tstrut{\rule{0pt}{2.6ex}}         
\newcommand\Bstrut{\rule[-0.9ex]{0pt}{0pt}}   

\ifCLASSINFOpdf
	 \usepackage[pdftex]{graphicx}
  	\graphicspath{Images/}
  	\DeclareGraphicsExtensions{.pdf,.jpeg,.png}
\else
\fi
\usepackage{amsmath}
\begin{document}
\bstctlcite{IEEEexample:BSTcontrol}
%
\title{A Recursive Method for Real-Time Waveform Fitting with Background Noise Rejection}
%
%

\author{A.~P.~Jezghani,
        L.~J.~Broussard,
        and~C.~B.~Crawford
\thanks{A. P. Jezghani was with the Department of Physics and Astronomy, University of Kentucky, Lexington, KY, 40506 USA. He is now with the Partnership for an Advanced Computing Environment, Georgia Institute of Technology, Atlanta, GA, 30332 USA e-mail:ajezghani3@gatech.edu}
\thanks{L. J. Broussard is with Oak Ridge National Laboratory, Oak Ridge, TN, 37830 USA.}
\thanks{C. B. Crawford is with the Department of Physics and Astronomy, University of Kentucky, Lexington, KY, 40506 USA}%
\thanks{This work is supported in part by the U.S. Department of Energy, Office of Nuclear Physics under Award Numbers DE-SC0008107, DE-SC0014622, and DE-AC05-00OR22725, and the National Science Foundation under Award No: PHY-0855584.}%
\thanks{L. J. Broussard was also supported by the Laboratory Directed Research and Development Program [project 8512] of Oak Ridge National Laboratory, managed by UT-Battelle, LLC, for the U.S. Department of Energy.}}
\maketitle

\begin{abstract}
We present here a technique for developing a high-throughput algorithm to fit a combination of template pulse shapes while simultaneously subtracting parameterized background noise. By convolving the psuedoinverse of the least-squares fit design matrix along a regularly sampled waveform trace, the time evolution of the fit parameters for each basis function can be determined in real-time. We approximate these sliding linear fit response functions using piecewise polynomials, and develop an FPGA-friendly algorithm to be implemented in high sample-rate data acquisition systems. This is a robust universal filter that compares well to common filters optimized for energy calibration/resolution, as well as filters optimized for timing performance, even when significant noise components are present.
\end{abstract}


%
\IEEEpeerreviewmaketitle

\section{Introduction}
%
%
%
%
\IEEEPARstart{D}{etection} of ionizing radiation is ubiquitous in science and technology, with applications ranging from medicine to security to fundamental science. These applications require various information about individual particles, such as detection time, deposited energy, and particle trajectory, or about the collection of radiation, like count rates, beam position, and focal spot size, all of which can be difficult to accurately determine in the presence of large noise. Advances in digital electronics have allowed for the migration from analog filters to digital pulse processing, resulting in greater overall flexibility, reproducibility, and sophistication in real-time analysis capabilities. In particular, field programmable gate arrays (FPGAs) have become a common platform for digital signal processing, aided especially by the development of several high-level synthesis options \cite{8356004}. 
With the departure from rigid analog shaping circuitry to  reconfigurable digital logic, powerful algorithms become available for agile response to changing conditions such as noise level or signal shape while maintaining detection accuracy.
In this paper, we present a universal framework in which basis functions representing both signal and noise contributions lead to finite impulse response (FIR) filters approximated by piecewise polynomials, and show how to implement them using minimal resources on FPGA.  These can rapidly be adapted to the current system state through configuration registers without having to re-synthesize the logic.

Numerous application-specific, FPGA-friendly FIR filters have been developed for the determination of energy and incident time $T_{0}$ for particle detectors, and are well-documented in literature \cite{nakhostin,KAMLEITNER201488}. For semiconductor detectors in particular, semi-gaussian, flat-top trapezoid, or cusp-like filters are commonly used due to their ability to improve signal-to-noise for step-like systems, restore exponentially decaying baselines, and integrate charge with ballistic deficit immunity \cite{JORDANOV1994337,TAN,ABBIATI,AGOSTINI}. For these same reasons, short implementations of these same filters are often implemented in parallel to serve as timing and pile-up filters \cite{XIA}. Other common filters for timing are based on differentiation to emphasize the step associated with particle arrival time, including constant fraction discriminators \cite{FALLU}, gaussian-smoothed derivatives \cite{YOUNG}, and RC-(CR)$^{n}$ filters with $n{\geq}1$ \cite{TINTORI}.

A challenge of any precision spectroscopy system is the rejection of noise superimposed on the detector response signal. Noise features can be correlated in time with the signal of interest, such as via stray capacitance and feedback in the amplification electronics, or uncorrelated, such as slow, transient baseline oscillations from microphonics of externally induced mechanical vibrations of detector electrodes or of electronics in a magnetic field. Many filters can attenuate noisy frequencies relative to the desired signal, but the resultant energy and timing resolutions are usually degraded, and the detection efficiency suffers.

A common technique to extract the desired signal is to perform a least squares fit of a template waveform alongside several basis functions that model the noise contributions. An important limitation for such a technique is the resource requirements for FPGA implementation; even by current standards, a traditional least squares fit requires too many resources to run in real-time with the acquisition. Efforts to develop and implement real-time least-squares type algorithms include a weighted least-squares estimator, with a more efficient diagonalized adaptation \cite{PETRICK92,PETRICK94,RIPAMONTI,JOLY}. However, these efforts lacked architecture-specific algorithm design or relied on input from additional filters. More recently, another method that has been developed is a recursive implementation, in which a highly down-sampled data stream is fit iteratively to determine the noise characteristics using 100s or 1000s of parameters \cite{474472,ZIMMERMANN2013404,7097420}. The fitted noise is then subtracted from the original, full sample-rate signal, which is then analyzed using more traditional techniques. Limitations of these techniques include reduced adaptability to changes in the noise behavior,  resulting in trigger efficiency issues, and high parameterization demands, which impede implementation on low-power FPGAs.  The method presented herein extends techniques used to develop the ubiquitous trapezoid filter~\cite{JORDANOV1994337} into a full implementation of least-squares fitter response functions using minimal resources which can be reconfigured on the fly.  This method is presented in section II, followed by a comparison of least squares filters with other common filters in section III.  The implementation on a user-configurable commercial digitizer is presented in section IV, followed by a discussion and conclusion in sections V and VI.

\section{Algorithm Design} \label{developing}
We describe development of the sliding least squares fitting algorithm in two independent steps: first the mathematical basis of generalized linear least squares fits with one nonlinear parameter ($T_0$), which can be implemented in a variety of contexts; and second, a pipelined implementation of arbitrary response functions, including the sliding least squares response, as a recursive piecewise polynomial FIR filter.

\subsection{Sliding Least Squares Fitter}

The sliding least squares fitter is an adaptation of the well-known generalized least squares fit \cite[sec. 15.4]{10.5555/1403886} to include the nonlinear parameter $T_0$ via convolution of individual response functions.  Any discretized waveform trace with samples $v[n]$, a column vector \textbf{v}, lies in an $N$-dimensional linear space and can be decomposed into the linear combination of $M$ basis functions, discarding uncharacterized random noise components. We describe this via the product of a design matrix, \textbf{A} (where each of the $M$  columns represents the $N$-sample time evolution of one basis function) and a column vector \textbf{a} (composed of the $M$ fit parameters). The basis functions are chosen from template waveforms which capture the shape of detection pulse signals, as well as noise patterns which should be excluded, such as slow baseline drifts or oscillations at specific frequencies. Multiplying the pseudoinverse $\mathbf{A}^+=(\mathbf{A}^{\mathrm{T}}\mathbf{A})^{-1}\mathbf{A}^{\mathrm{T}}$ of the design matrix with the waveform trace \textbf{v} yields the solution to the least-squares fit problem for fixed-time basis functions,
\begin{equation}
\mathbf{a}=\mathbf{A^{+}}\mathbf{v},
\end{equation}
where the minimum chi-square at this optimal value of $\mathbf{a}$ is
\begin{equation}
\chi^2=\mathbf{v}^{\mathrm{T}}(\mathbf{I}-\mathbf{AA}^+)\mathbf{v}=\mathbf{v}^{\mathrm{T}}\mathbf{v}-\mathbf{a}^{\mathrm{T}}\mathbf{B}\,\mathbf{a},
\end{equation}
where $\mathbf{B}=\mathbf{A}^{\mathrm{T}} \mathbf{A}$ is the inverse covariance matrix of $\mathbf{a}$, assuming $\mathbf{v}$ has been normalized by its standard deviation $\sigma_v$.

In the description so far, the time offset of each basis function is fixed, and can only fit one $N$-sample segment of the continuous waveform at a time, which would require preknowledge of the true pulse start time $T_0$.  However, for a waveform much longer than the length of the design matrix, the convolution of each row of the pseudoinverse of \textbf{A} with \textbf{v} yields the time evolution of the vector of fit parameters, \textbf{a} as a function of the start time $n$:
\begin{equation}
\mathbf{a}[n]=\mathbf{A^{+}} \boldsymbol{*} {v}[n],
\end{equation}
where $\boldsymbol{*}$ represents the vector of $M$ independent convolutions ${a_m}[n]=\mathbf{A}_m^{+} * {v}[n]$ of the $m^\mathrm{th}$ row of $\mathbf{A\textsuperscript{+}}$ with $\mathbf{v}$ to obtain the $m^\mathrm{th}$ component of $\mathbf{a}$ as a function of $n$, for $m=1,2,\ldots M$. On CPU, and especially GPU, this can trivially be implemented as an FFT convolution. By minimizing
\begin{equation}
\chi^2[n]=\mathbf{1} * v[n]^2 -\mathbf{a}[n]^{\mathrm{T}}\mathbf{B}\,\mathbf{a}[n],
\end{equation}
as a function of $n$, (where $\mathbf{1}$ is a vector the same length as $\mathbf{A}_m^+$ but with all components equal to 1,  $v[n]^2$ is squared pointwise, and the matrix sandwich product $\mathbf{a}^{\mathrm{T}}\mathbf{B}\,\mathbf{a}$ is evaluated at each point $n$), the non-linear fit parameter $T_{0}$ can be determined from the start sample $n_{\mathrm{min}}$ with minimal $\chi^2$.  The optimal linear fit parameters are then $\mathbf{a}[n_{\mathrm{min}}]$.

As a practical matter, it is convenient, more numerically stable, and often just as accurate to find $n_{\mathrm{min}}$ by maximizing the combination $r_t[n]=\mathbf{b}^{\mathrm{T}}\mathbf{a}[n]$ of the fit parameters, for a fixed vector $\mathbf{b}$ describing the relative energy of each of the $M$ pulse templates in $\textbf{A}$, with $b_m=0$ for each noise basis function.  In the trivial case of one template waveform, $r_t[n]$ reduces to the corresponding component of $\mathbf{a}[n]$. 

\subsection{Recursive Piecewise Polynomial Filter}

To optimize the use of the finite resources on the FPGA, filter logic used for convolutions in the sliding least squares fitter must be lightweight and limited to simple logic like fixed-point addition and multiplication, delays, registers, and accumulators.  Furthermore, filters which output one new filtered response point per clock cycle are desirable on digitizers to process the data in real time as they are acquired, for example, in threshold triggers. These two requirements preclude both the simplistic sliding vector inner product, which requires $n$ multiplies per clock cycle, and FFT-based convolutions, which process waveforms in batches. It is more efficient to chain the output of multiple sequential, recursive polynomial filters of different lengths, one after another.  Such a recursive piecewise polynomial filter can approximate the response function of arbitrarily shaped kernels, significantly reducing the computational complexity.  This section describes the recursive implementation of one segment 
\begin{equation}
\label{eq:taylor}
h[n]=c_0+c_1 n+c_2 n^2+\ldots+c_K n^K,\quad 1\leq n \leq L
\end{equation}
of a piecewise polynomial impulse response function $h_t[n]$ of length $L$ and order $K$ which, for example, can approximate the combination $\mathbf{b}^{\mathrm{T}}\mathbf{A}^+$ of filters used in the sliding least squares fitter.  The sum of the convolutions with each polynomial segment of the form 
\begin{equation}
    r[n]=h[n]*v[n] = \sum_{n'=1}^L h[n']\,v[n-n'+1],
\end{equation}
with each segment appropriately delayed, yields the time-series of the corresponding combination of fit parameters $r_t[n]=\mathbf{b}^{\mathrm{T}}\mathbf{a}[n]$. 

\begin{figure}[!ht]
\centering
\includegraphics[width=2in]{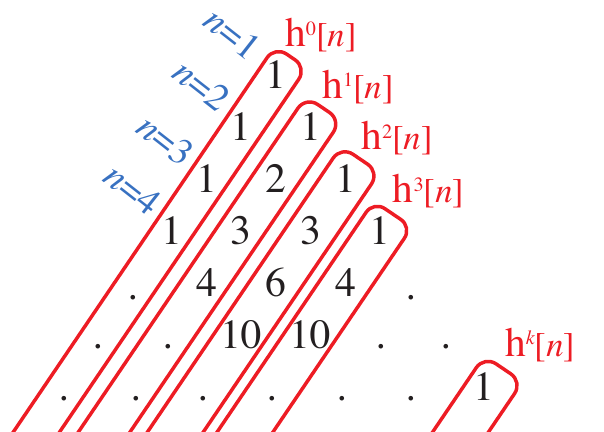}
\caption{Impulse response functions $h^k[n]$ as diagonals of Pascal's triangle.}
\label{fig:rec_poly_pascal}
\end{figure}

\begin{figure}[!ht]
\centering
\includegraphics[width=3.4in]{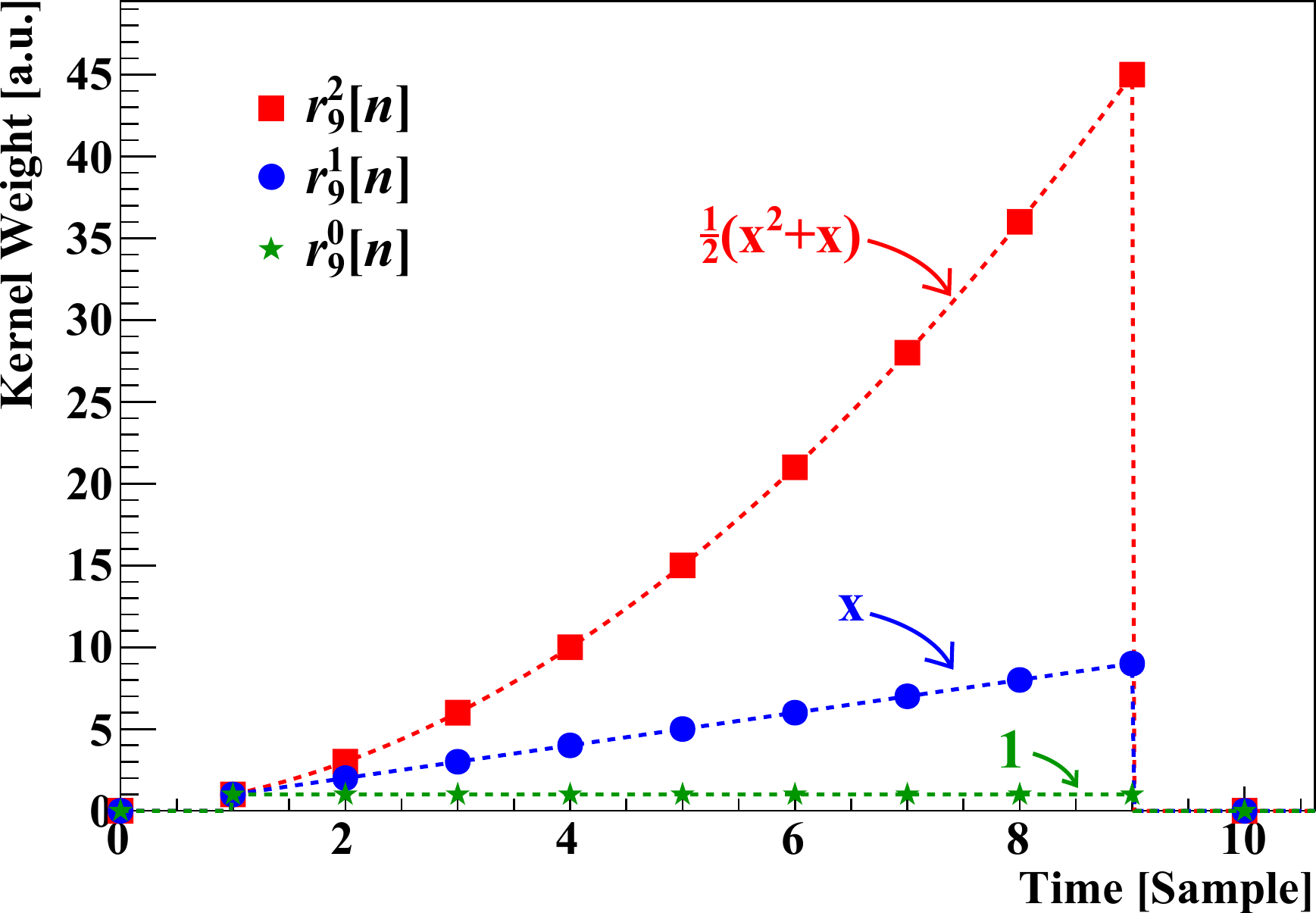}
\caption{Impulse response functions for recursive polynomial convolutions of length 9, and their corresponding Pochhammer polynomials (dashed lines).  The coefficients of each polynomial are given by the columns of Eq.~\ref{eq:coef_trans}.}
\label{fig:rec_poly_kern}
\end{figure}

As demonstrated to second order by Jordanov and Knoll~\cite{JORDANOV1994337}, finite length polynomial impulse response functions can easily be achieved via repeated accumulation followed by delayed subtraction at each step to truncate the response to finite length $L$. Heuristically, each accumulator yields a higher order polynomial, resulting in the truncated impulse response $h_L^{k}[n]$ of order $k$ after $k+1$ integrations. Due to the discrete summations in these convolutions, the untruncated response is not a simple power of $n$, but instead the polynomial
\begin{equation}
h^{k}[n]=\frac{n^{(k)}}{k!}=\binom{n+k-1}{k},
\end{equation}
where $x^{(k)}$ is the Pochhammer polynomial of order $k$,
\begin{equation}
\label{eq:pochammer}
x^{(k)}=\sum_{j=0}^k |S_{k}^{(j)}| x^j = \left\{ \,
\begin{IEEEeqnarraybox}[][c]{l?s}
\IEEEstrut
1 & if $k=0$, \\
 \prod_{j=0}^{k-1} (x+j)& if $k\geq1$,
\IEEEstrut
\end{IEEEeqnarraybox}
\right.
\end{equation}
and $|S_{k}^{(j)}|$ are unsigned Stirling numbers of the first kind~\cite[sec. 24.1.3]{AbramowitzStegun}. 
The vectors $\mathbf{h}^k$ form the diagonals of Pascal's triangle (Fig.~\ref{fig:rec_poly_pascal}), as apparent from the corresponding recursion relation of binomial coefficients.  The first three response functions of length $L=9$ are plotted in Fig.~\ref{fig:rec_poly_kern}.


\begin{figure*}[!hbt]
\centering
\includegraphics[width=\textwidth]{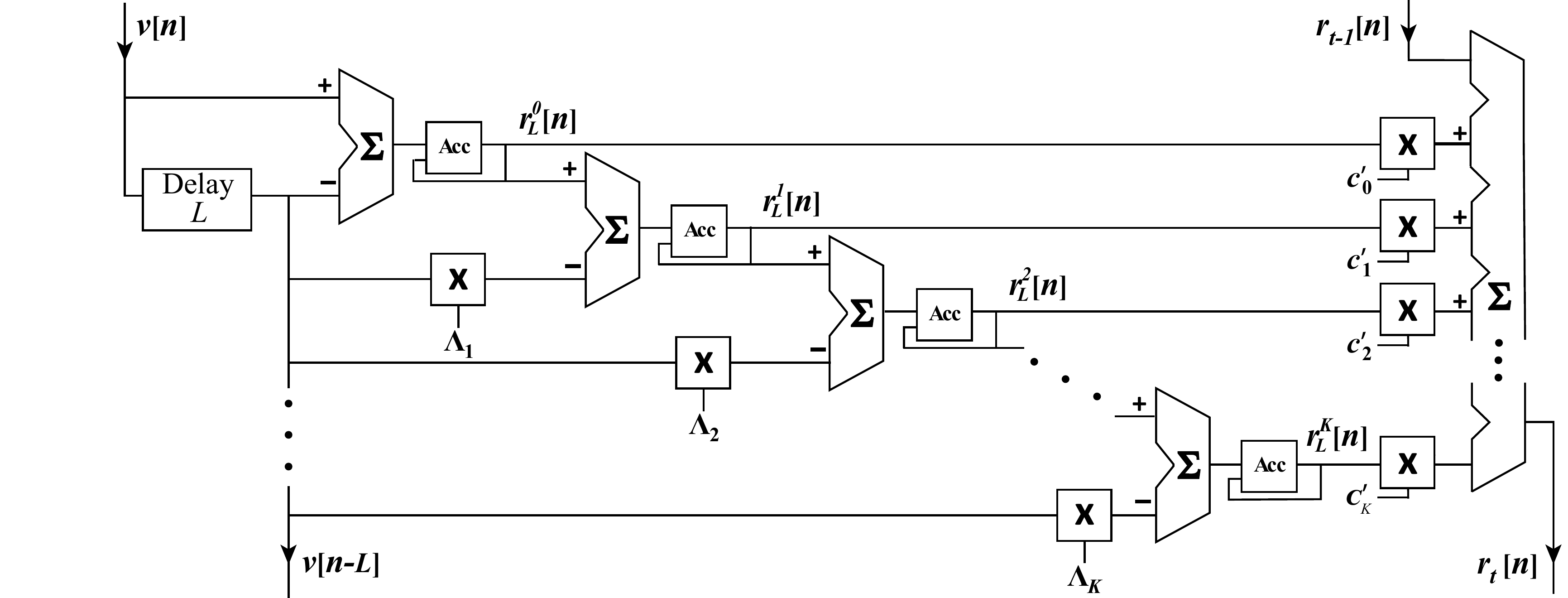}
\caption{Logic diagram for recursive implementation of the convolution $r[n]=h[n]*v[n]$ of one segment (Eq.~\ref{eq:taylor}) of the piecewise polynomial impulse response $h_t[n]$ with the input signal $v[n]$, adding the result $r[n]$ to the cumulative output $r_t[n]$. The configuration coefficients $\Lambda_k$ and $c'_k$ are obtained from $c_j$ via Eqs.~\ref{eq:lambda} and \ref{eq:cprime}. Sequential segments of the recursive piecewise polynomial filter are chained vertically, with the input $v[n]$ on the top left of the first segment, and filtered output $r_t[n]$ on the bottom right of the last segment.}
\label{fig:rec_poly_logic}
\end{figure*}

Extending the results of~\cite{JORDANOV1994337}, the $k+1$ serial integrations and truncations of the output $r^k_L[n]=h^k_L[n]*v[n]$ are
\begin{IEEEeqnarray}{rCl}
\label{eq:ppsumm}
r^{k}_{L}[n] & = & \sum_{i_{k}=0}^{n}\left( \sum_{i_{k-1}=0}^{i_{k}}\left(\ldots  \sum_{i_{0}=0}^{i_{1}}\bigg(v[i_{0}]-\Lambda_0 v[i_{0}-L]\bigg)\ldots\right.\right. \nonumber\\
&&\left.\vphantom{\sum_{i_{k-1}=0}^{i_{k}}}\left.\vphantom{\ldots  \sum_{i_{0}=0}^{i_{1}}} -\Lambda_{k-1} v[i_{k-1}-L]\right) - \Lambda_{k} v[i_{k}-L]\right),
\end{IEEEeqnarray}
where each delayed subtraction involving the total integral
\begin{equation}
\label{eq:lambda}
    \Lambda_{k}\equiv h^k[L] =\sum_{n=1}^{L\,\mathrm{or}\, \infty}{h_L ^{k-1}[n]}, \quad \Lambda_0=1
\end{equation}
truncates $h_L^k[n]$ to zero for all $n>L$.  Defining $r_L^{-1}[n]\equiv v[n]$, 
Equation \ref{eq:ppsumm} has the explicit recursive representation
\begin{equation}
    \label{eq:rec}
    r^{k}_{L}[n]=r^{k}_{L}[n-1]+r^{k-1}_{L}[n]-\Lambda_{k}v[n-L],
\end{equation}
which is used to wire the digital recursive filter of Fig.~\ref{fig:rec_poly_logic}.  $r_l^0[n]$, $r_{k'}^1[n]$, and $r_k^2[n]$ correspond to $p(n)$, $r(n)$, and $u(n)$, respectively, of Equations 22--25 in reference~\cite{JORDANOV1994337}. Fig.~\ref{fig:rec_poly_kern} shows the relationship between the recursive polynomial responses and their continuous time counterparts.

Each polynomial response segment of individual order $K$ and length $L$ may be expanded in the above basis of naturally synthesizable polynomials as
\begin{equation}
    h[n]=\sum_{k=0}^K c'_k h_L^k[n] 
    = \sum_{j=0}^K \left( \sum_{k=j}^{K}\frac{|S^{(j)}_k|}{k!}c'_k\right) n^j
\end{equation}
where the original coefficients $c_j$ of Equation~\ref{eq:taylor} are given in parentheses.  The inverse linear transformation is
\begin{equation}
\label{eq:coef_trans}
\begin{pmatrix}
c'_{0}\vphantom{|S_{K}^{(0)}|/K!} \smallskip\\
c'_{1}\vphantom{|S_{K}^{(1)}|/K!}\smallskip\\
c'_{2}\vphantom{|S_{K}^{(2)}|/K!}\smallskip\\
c'_{3}\vphantom{|S_{K}^{(3)}|/K!}\\
\vdots\\
c'_{K}\vphantom{|S_{K}^{(K)}|/K!}
\end{pmatrix}
=
\begin{pmatrix}
1         & 0         & 0         &  0      & \hdots & 0\vphantom{|S_{K}^{(0)}|/K!} \smallskip\\
0         & 1         & 1/2      & 1/3      & \hdots & |S_{K}^{(1)}|/K!\smallskip\\
0         & 0         & 1/2      & 1/2      & \hdots & |S_{K}^{(2)}|/K!\smallskip\\
0         & 0         & 0        & 1/6       & \hdots& |S_{K}^{(3)}|/K!\smallskip\\
\vdots & \vdots & \vdots & \vdots & \ddots & \vdots \\
0         & 0         & 0        & 0        & \hdots & |S_{K}^{(K)}|/K!
\end{pmatrix}^{\!\!\!-1}
\begin{pmatrix}
c_{0}\vphantom{|S_{K}^{(0)}|/K!}\smallskip\\
c_{1}\vphantom{|S_{K}^{(1)}|/K!}\smallskip\\
c_{2}\vphantom{|S_{K}^{(2)}|/K!}\smallskip\\
c_{3}\vphantom{|S_{K}^{(3)}|/K!}\\
\vdots \\
c_{K}\vphantom{|S_{K}^{(K)}|/K!}
\end{pmatrix}.
\end{equation}

The $c'_k$ are obtained recursively by back substitution,
\begin{equation}
\label{eq:cprime}
c'_K=c_K K!, \quad c'_{j-1}=\left(c_{j-1}-\sum_{k=j}^{K}\frac{|S^{(j-1)}_k|}{k!}c'_k\right) j!.
\end{equation}

Figure \ref{fig:rec_poly_logic} depicts the logic diagram of one recursive polynomial convolution segment $r[n]=\sum_{k=1}^K c'_k r^k_L[n]$, patterned after Equations~\ref{eq:rec}, where $c'_k$ are multiplied by the respective output stage to produce the full convolution with $h[n]$. Successive pipelining of the delayed original signal to each proceeding segment produces the full piecewise polynomial convolution $r_t[n]$.  

The original coefficients $c_j$ of each segment are obtained by fitting the ideal response function to a piecewise polynomial with segments described by Eq.~\ref{eq:taylor}, where the order $K$ and length $L$ of each segment may be fixed or optimized as part of the fitting process.  Because $h_L^k[n']$ starts back at $n'=1$ for each segment, the coefficients $c_j$ must be fit to polynomials in the offset time $n'$, not the global time $n$ of $h_t[n]$. In this work, we placed knots between each polynomial segment at the local extrema when fitting a piecewise polynomial to the ideal response functions, but more sophisticated techniques can be used to reduce the polynomial complexity~\cite{BUTLER20114076}. Each segment of the kernel was fit to the lowest order polynomial such that $\chi^2$ was less than a specified amount.

\begin{figure}[!ht]
\centering
\includegraphics[width=3.3in]{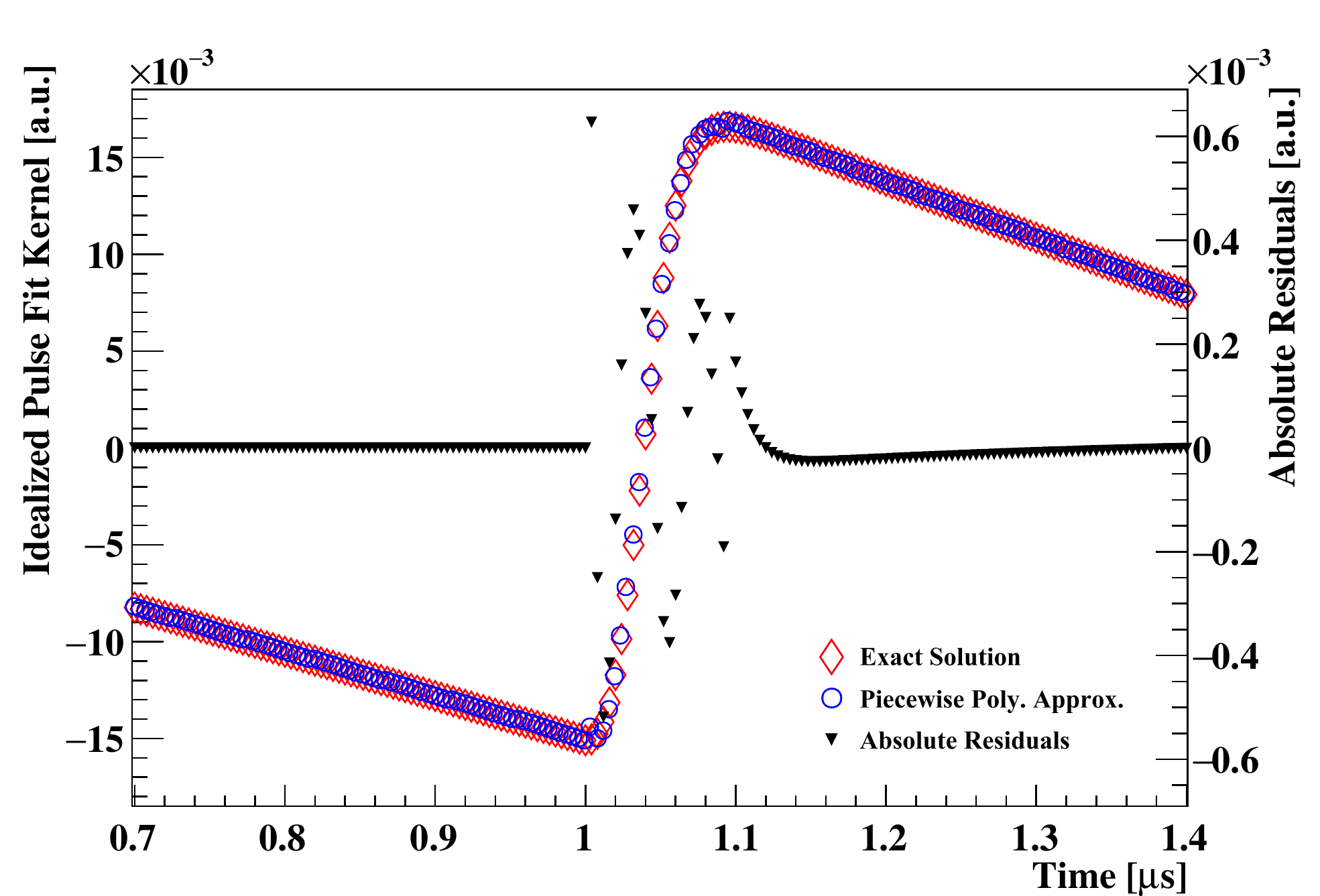}
\caption{A least squares fit kernel based on an idealized tail pulse shape, and a 2nd-order truncated series expansion for an oscillating baseline. The red open diamond and blue open circle markers indicate the kernel weight evolution as a function of time for the analytic and approximated expressions, and the black triangle markers show the residual difference between the two.}
\label{fig:pulse_fit_kern}
\end{figure}

One final consideration for FPGA implementation is the conversion of scalar multiplication from floating point to fixed point arithmetic, the natural data type for real numbers on FGPAs. This will introduce  minor rounding error depending on the precision of allocated resources, which, when combined with the piecewise polynomial approximation, can result in deviations from the true convolution, such as a non-zero integral area that would develop filter instability. A simple yet effective solution is to modify the constant coefficient to correct for this integral offset, and restore the zero area property. Figure \ref{fig:pulse_fit_kern} shows a comparison between the true pseudoinverse fit vector for an idealized tail pulse with oscillating background compared to an FPGA-appropriate recursive piecewise polynomial approximation. As expected, the greatest deviation between the two occurs in regions with large 2\textsuperscript{nd} derivative, where the fitted polynomial lags slightly behind the ideal response function. 
These deviations were sufficiently small to have negligible impact on the final performance of the filter in the tests reported herein, but the fit can be improved as required for more stringent applications.

\section{Filter Characterization}
\subsection{Test Data Set}
\begin{figure}[!ht]
\centering
\includegraphics[width=3.3in]{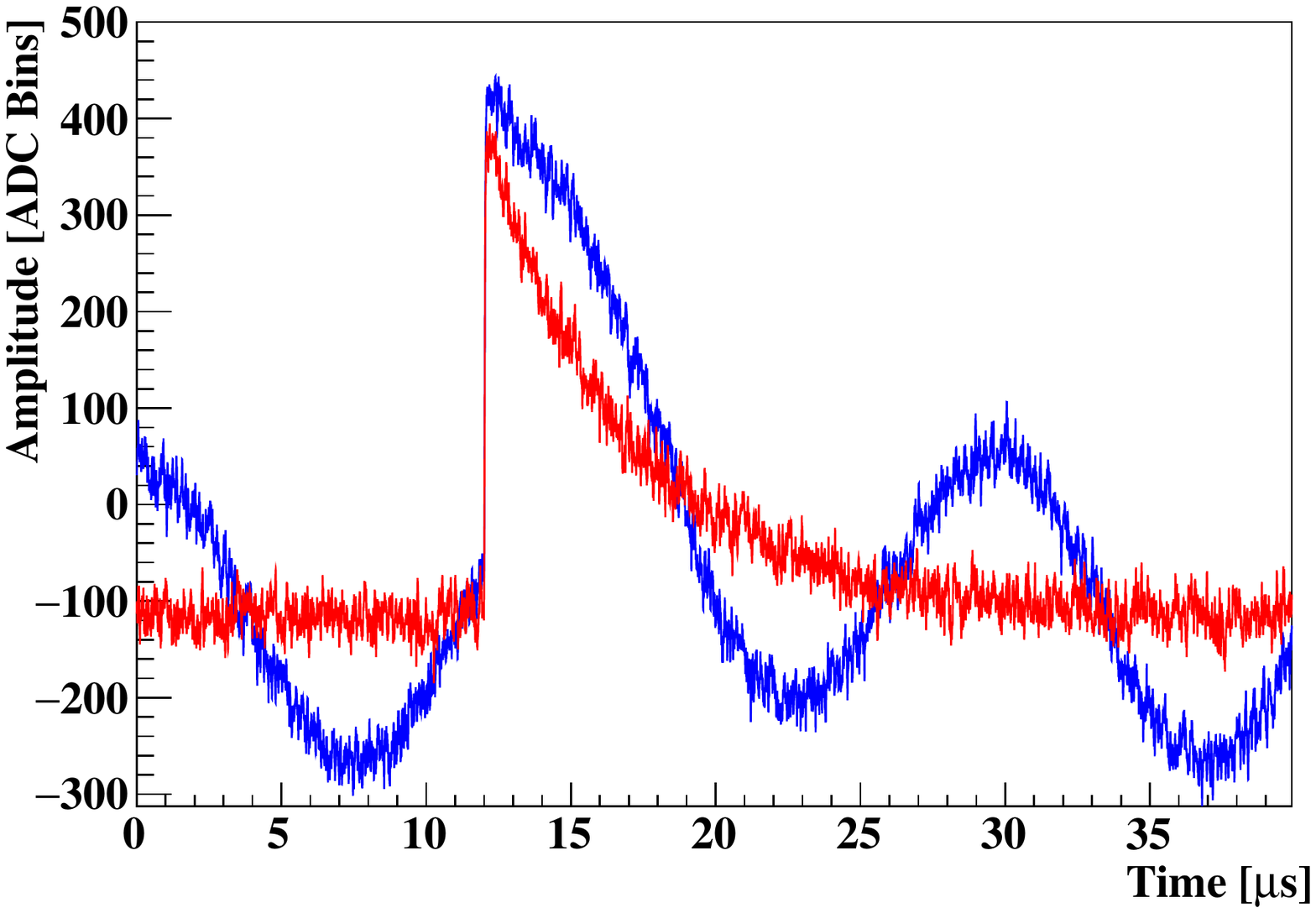}
\caption{Sample synthetic waveform traces. Each has baseline noise with an RMS width of approximately 18 ADC bins and an offset of -125 ADC bins. The slow oscillations have an amplitude of 150 ADC bins, and vary in frequency and phase for each event. The pulse shaping characteristics reflect those of the Nab silicon detection system.}
\label{fig:samp_synth_wf}
\end{figure}

In order to compare the performance of the different filtering algorithms, a synthetic data set was constructed based on experimental data observed in the prototype detection system for the Nab neutron beta decay correlation experiment \cite{BROUSSARD201783}. Each simulated waveform was generated as an integrated semi-gaussian current pulse with variable rise time to mimic charge collection effects in the silicon, followed by the convolution with a CR-(RC)\textsuperscript{2} shaper to emulate the front-end electronics \cite{jezghani}. Random noise, generated from an empirically determined average power spectrum with uncorrelated phases, was then superimposed on top of the idealized pulse shape. Additionally, random-phase oscillations with frequencies ranging from 25-80 kHz and amplitudes of 150 ADC bins were added to mimic the transient microphonics noise observed in the data set. For the study, the initial rise of each simulated pulse starts at the same time bin, and the amplitude spans from 20 to 5000 ADC bins, the anticipated range for events in the Nab experiment. Figure \ref{fig:samp_synth_wf} shows example synthetic pulses used in this analysis.

\begin{figure}[!ht]
\centering
\includegraphics[width=3.3in]{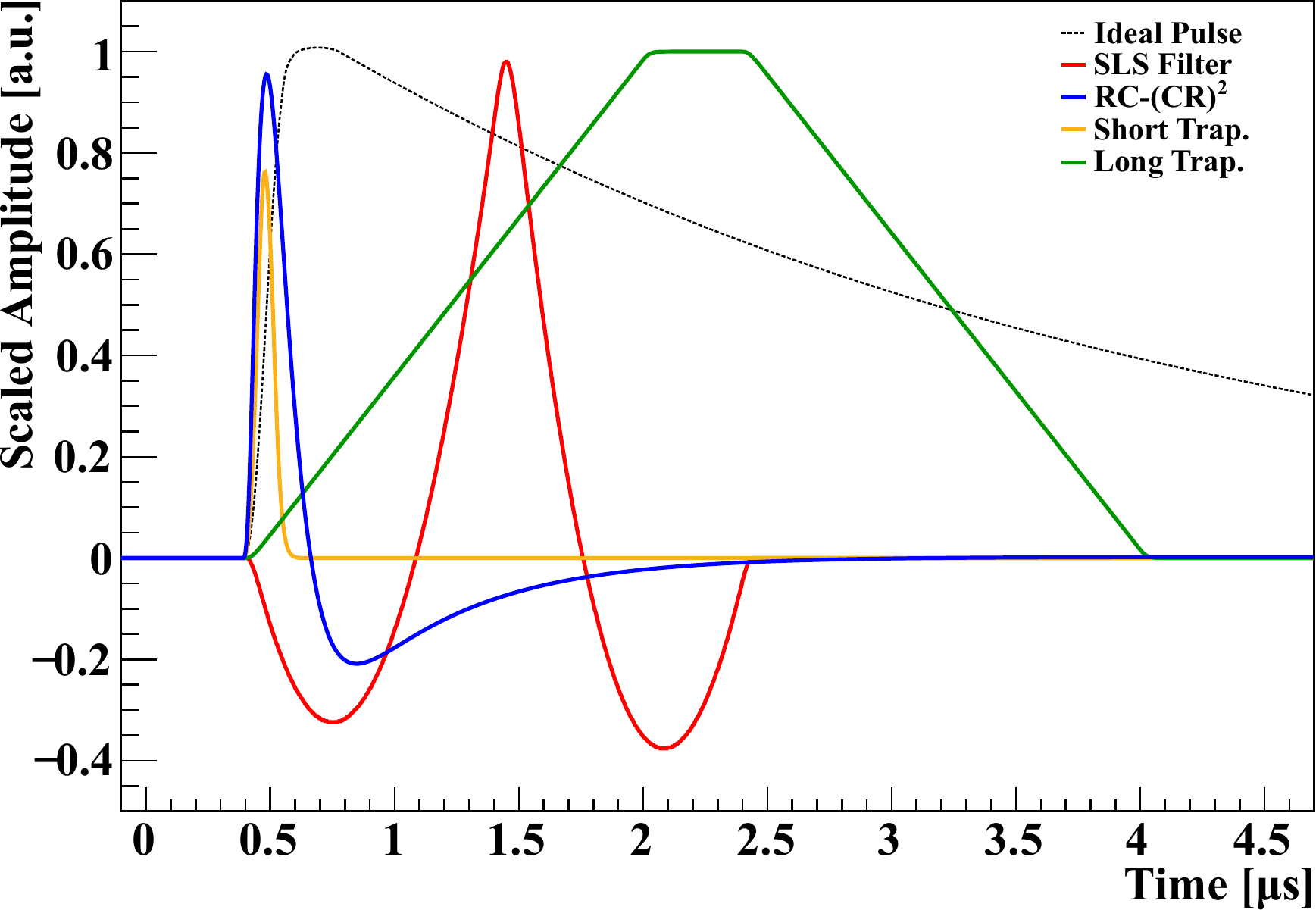}
\caption{The filter output after convolution with an ideal pulse for each of the filters used in the comparative analysis (color online).}
\label{fig:filters}
\end{figure}

The test data were analyzed using the sliding least squares fit convolution described in section \ref{developing}, as well as a long trapezoid (ideal for energy reconstruction), a short trapezoid (used for timing), and an RC-(CR)\textsuperscript{2} filter (also used for timing). The output of each filter on an ideal tail pulse is shown in figure \ref{fig:filters}. 

Each filter's performance was analyzed using several metrics including accuracy in energy reconstruction, accuracy and resolution in timing, and trigger efficiency, for relatively clean signals as well as signals with a random-phase low frequency baseline oscillation.

\begin{table*}[tbp]
\caption[]{Synthetic data analysis filter parameters.}
\centering
\begin{tabular}{| l | c | c | c |}
\hline
\textbf{Filter} & \textbf{Characteristics} & \textbf{Eff. Thresh. [ADC Bins]} & \textbf{Avg. No. False Triggers} \Tstrut\Bstrut\\
\hline
SLS Filter & \unit{2}{\micro\second} fit basis, uniform weighting & 44 & 52.8\Tstrut\Bstrut\\
\hline
Short Trap. & \unit{80}{\nano\second} rise time, \unit{0}{\nano\second} flat top & 87 & 49.6\Tstrut\Bstrut\\
\hline
RC-(CR)\textsuperscript{2} & \unit{512}{\nano\second} RC stage, \unit{64}{\nano\second} CR stages & 95 & 51.9\Tstrut\Bstrut\\
\hline
Long Trap. & \unit{1.6}{\micro\second} rise time, \unit{0.4}{\micro\second} flat top & 27 & 42.3 \Tstrut\Bstrut\\
\hline
\end{tabular}
\label{tab:filtpar}
\end{table*}

For each filter, the threshold was set such that approximately 0.5\% of the triggers generated were due to noise fluctuations. Table \ref{tab:filtpar} details the shaping parameters and threshold for each filter, as well as the average number of false triggers for the 10k waveform sets for each of the 15 amplitude settings. In the event of a trigger, the energy and time for the short trapezoid, sliding least squares fitter, and RC-(CR)\textsuperscript{2} were determined based on the value of the local maximum within a short window of the trigger. The long trapezoid time and energy were determined as the midpoint of the flat top region, as determined by a rising and falling edge trigger through a fixed threshold. Each trigger is followed by a fixed dead-time of \unit{5}{\micro\second} to prevent reduntant triggering on the same event.

\subsection{Energy Reconstruction}
\begin{figure}[!b]
\centering
\includegraphics[width=3.3in]{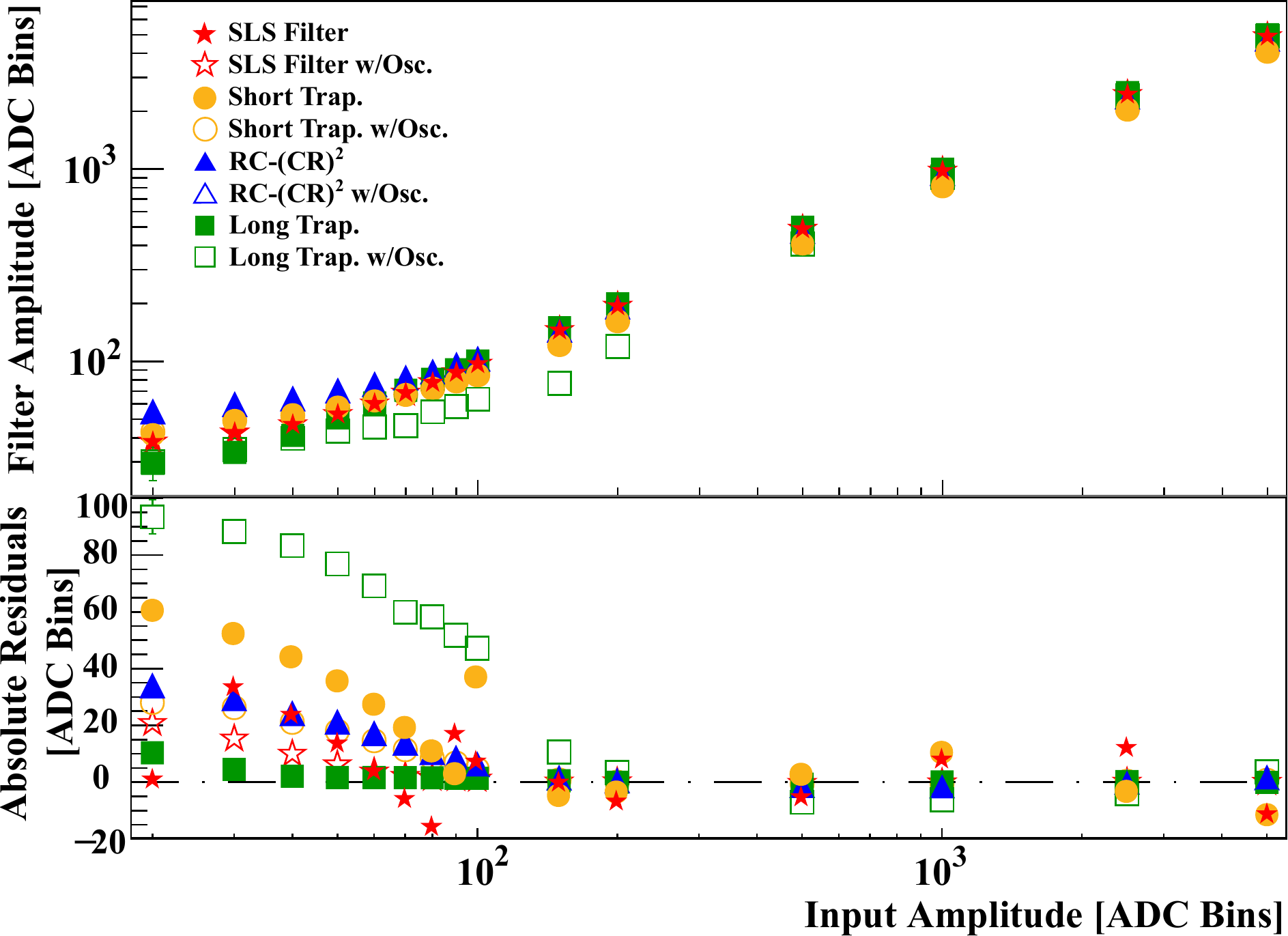}
\caption{The measured amplitude for each of the filters with and without noise (top), and the residuals to a linear fit for input amplitudes $>$150~ADC bins.}
\label{fig:ener}
\end{figure}

\begin{figure}[!b]
\centering
\includegraphics[width=3.3in]{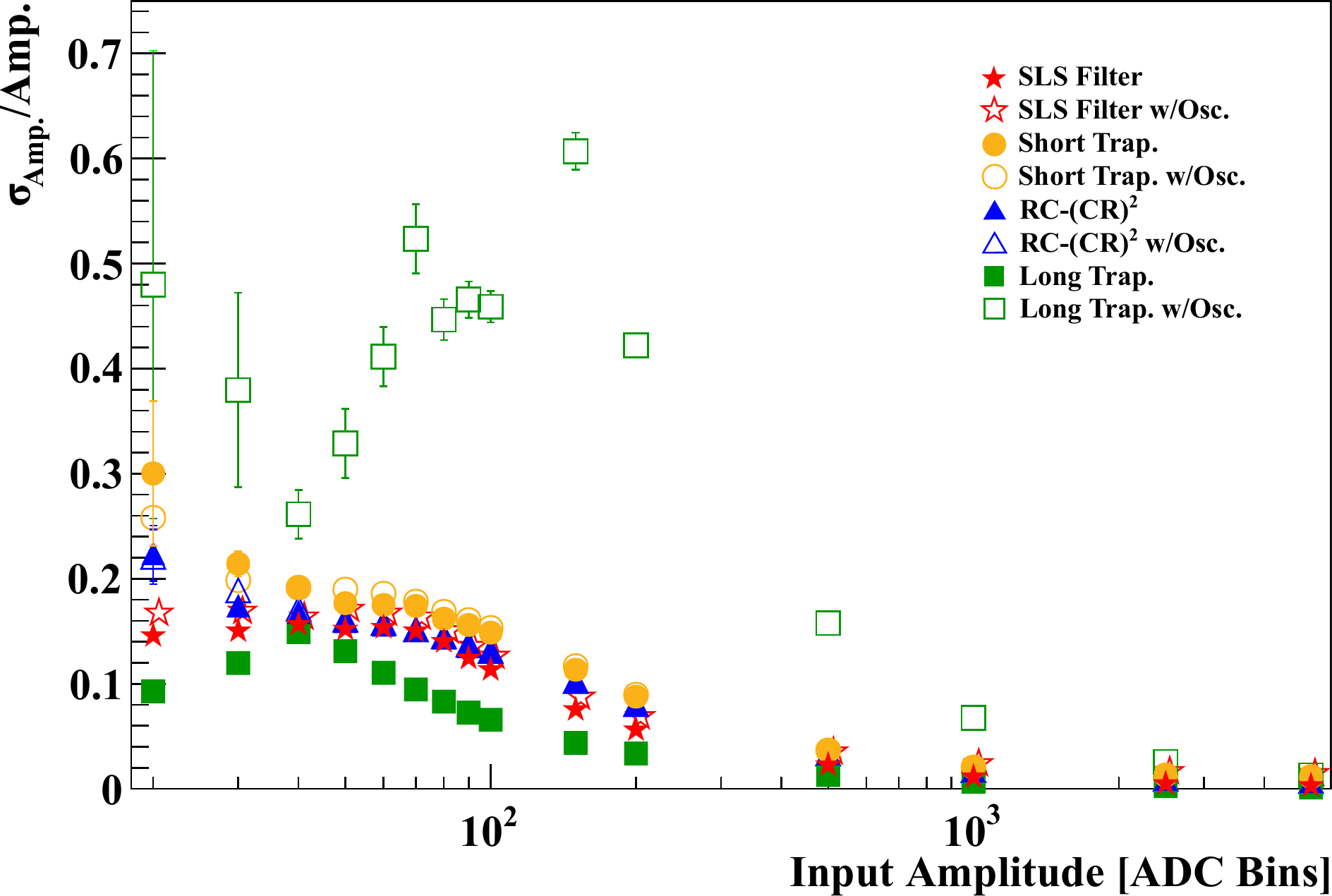}
\caption{The relative peak widths as determined by a gaussian fit.}
\label{fig:en_res}
\end{figure}

Figures \ref{fig:ener} and \ref{fig:en_res} show the extracted pulse amplitude and relative resolution for each of the different filters. As expected, the long trapezoid achieves the best results in the absence of baseline oscillations. The long shaping time effectively filters the high frequency noise, while the flat top properly integrates the charge while compensating for the exponential baseline. Furthermore, determining the energy from a fixed position on top of the trapezoid prevents a systematic bias that occurs when a local maximum is used instead. Nonetheless, the other three filters maintain linearity within $\pm15$~ADC bins for most of the range of interest.

With the addition of slow baseline oscillations, the short trapezoid,  RC-(CR)\textsuperscript{2}, and  least squares fitter perform comparably to the cleaner baseline. The two timing filters were unaffected thanks to their short lengths, while the fitter successfully fits to the ideal pulse shape while subtracting the oscillations. The long trapezoid's performance is degraded, particularly in the low amplitude regime, as the energy pick-off is determined by the rising and falling edge triggers, which are more difficult to determine with the additional noise features.

\subsection{Timing Resolution and Accuracy}
\begin{figure}[!ht]
\centering
\includegraphics[width=3.5in]{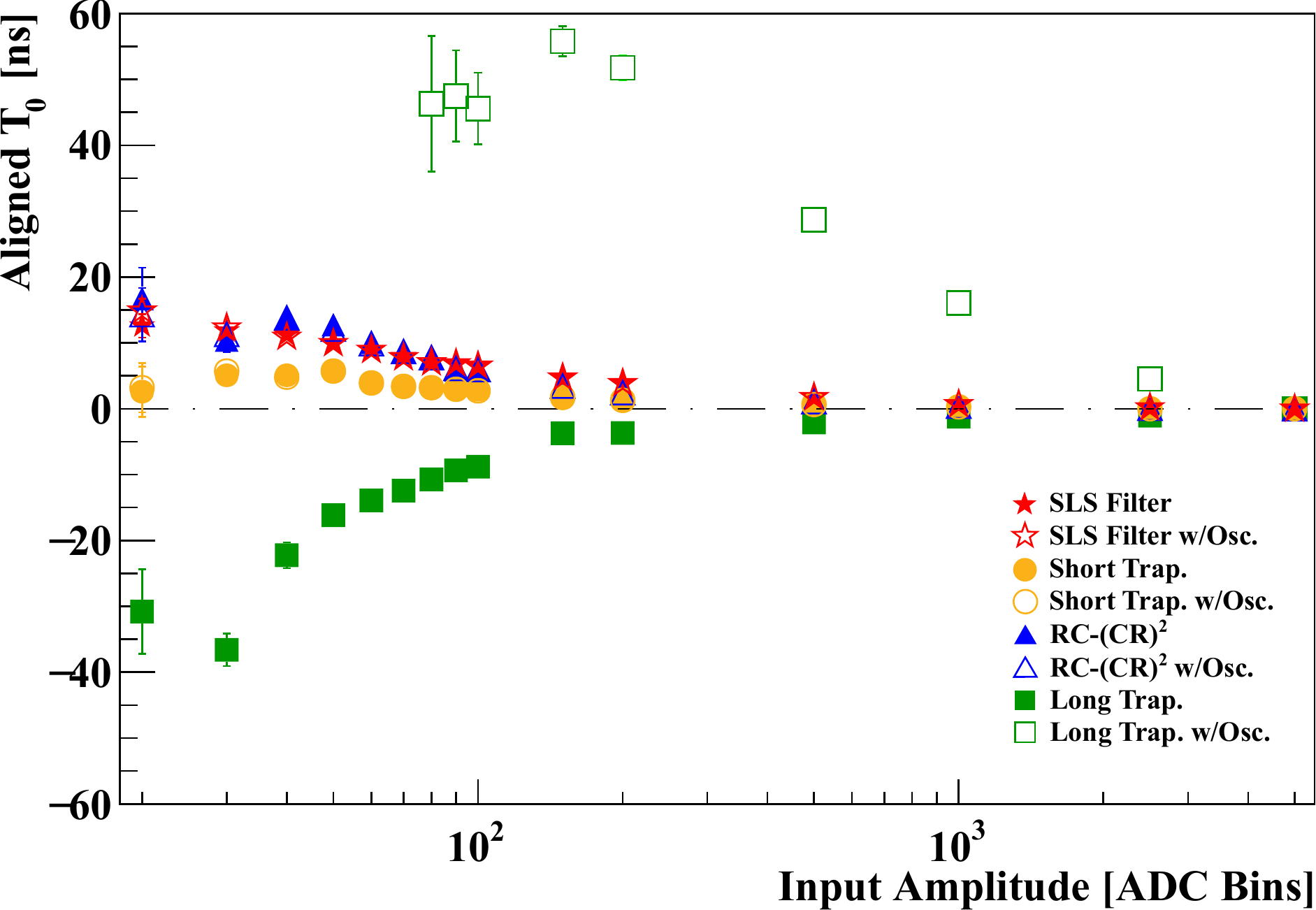}
\caption{The corrected trigger time, $T_{0}$, for each of the filters. The low amplitude data for the long trapezoid with the oscillating baseline is off scale at low amplitudes.}
\label{fig:t0}
\end{figure}

Each filter has an inherent offset from the start of the rising edge of a pulse due to the shaping characteristics of that filter. Figure \ref{fig:t0} shows the extracted $T_{0}$ for each filter, after correcting for this offset based on an ideal pulse response. A deviation in the corrected $T_{0}$ from 0 as a function of amplitude is indicative of a systematic bias in the results; as such, the ideal filter should minimize this drift. Both timing filters and the least squares fitter are reasonably robust against the presence of a bias, while the long trapezoid deviates more strongly with amplitude.

\begin{figure}[!ht]
\centering
\includegraphics[width=3.5in]{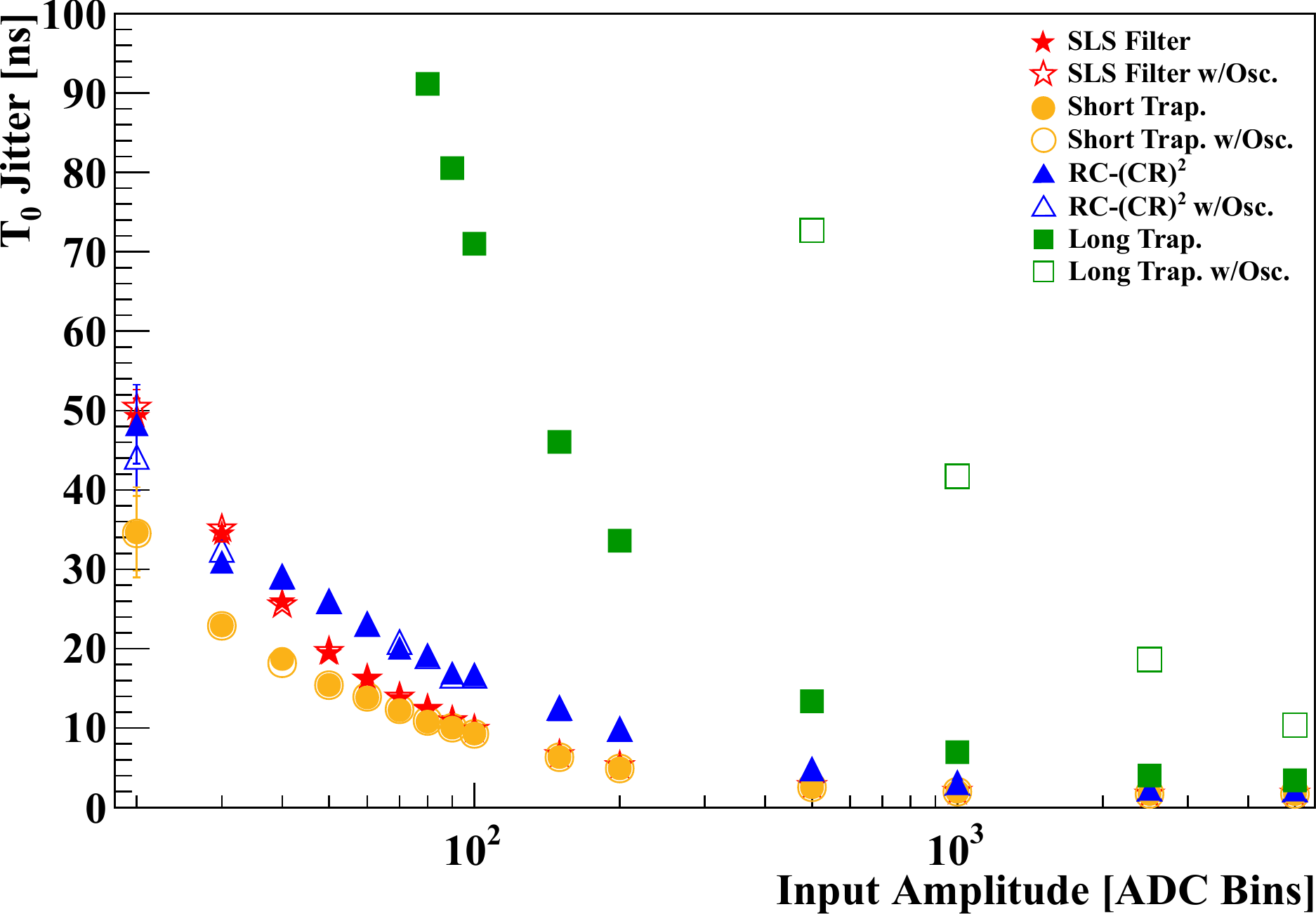}
\caption{The standard deviation of the trigger time distribution for each of the filters. The long trapezoid data runs off-scale at lower amplitudes.}
\label{fig:t0jitter}
\end{figure}

In addition to the accurate reconstruction of the event start time, another important consideration is the jitter in the determined $T_{0}$, which can be determined from the width of the trigger time distribution. While the start time of a pulse can be refined greatly with offline analysis, jitter in the online trigger can complicate this process, since the variation of the pulse relative to the pretrigger length necessitates additional parameterization. Figure \ref{fig:t0jitter} shows the standard deviation for each of the filters. The long trapezoid exhibits the greatest variation in trigger timing, both with and without the baseline oscillations.

\subsection{Trigger Efficiency}
\begin{figure}[!ht]
\centering
\includegraphics[width=3.5in]{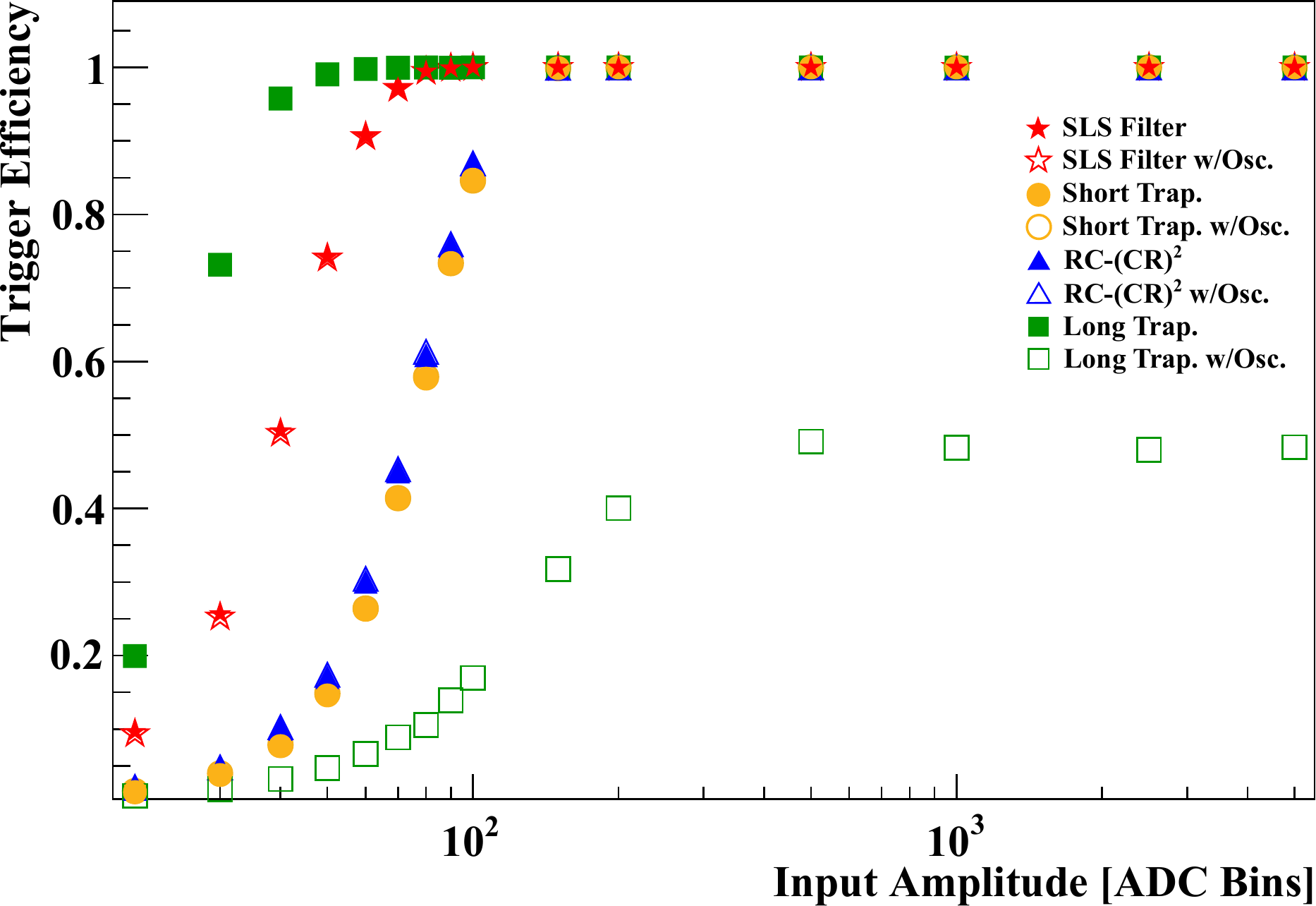}
\caption{The trigger efficiency as a function of pulse amplitude for each of the filters, with and without noise. The fast timing filters and least squares fitter are immune to the low frequency baseline oscillations, but the long trapezoid trigger efficiency suffers significantly with the additional noise.}
\label{fig:effic}
\end{figure}

Figure \ref{fig:effic} shows the fraction of observed triggers as a function of amplitude for each of the filters. With a stable baseline, the long trapezoid, with its long baseline averaging, is capable of the lowest threshold for high efficiency triggering. The two timing filters, however, require larger amplitude pulses for efficient detection. The least squares fitter, while not as efficient as the long trapezoid, still maintains better than 99.9\% efficiency for events with energy comparable to than that of a proton detected in the Nab experiment.

The RC-(CR)\textsuperscript{2} and short trapezoid perform comparably in the presence of slow baseline oscillations, as their short shaping times integrate little of the oscillation. The long trapezoid is much more susceptible to the baseline fluctuations, and the trigger efficiency suffers dramatically, even at large amplitudes. The least squares fitter fits for the ideal pulse response while subtracting the slow oscillation, and as such preserves its low-threshold trigger efficiency even with the noise.

\section{FPGA Implementation}
As a proof of concept, we implemented a generic recursive piecewise polynomial filter on the NI PXIe-5171R digitizer. Each module has eight channels of 250 MS/s, 14-bit ADCs backed by a Kintex-7 410T FPGA. The FPGA is user-programmable using the FPGA module of the graphical LabVIEW language.  The FPGA reads two samples from each ADC at 125 MHz, which are pairwise averaged and feed directly into  the online trigger logic. Each channel is processed by a filter comprised of seven segments, each with maximum length of 500 samples (\unit{4}{\micro\second}), of fourth order polynomials. The polynomial coefficients, $c'_{k}$, and length parameters, $\Lambda_{k}$, for each channel and segment are passed to the FPGA at run-time, allowing for iterative filter development without the added cost of firmware recompilation. Each of the coefficients is implemented as a 55-bit fixed point number, with 50 fractional bits, and all compression is delayed until the final filter output. Implementation of the filter logic in look-up tables and shift registers (as opposed to DSP slices) consumes an additional 5.6\% and 4.5\% of the resources, respectively.

\section{Discussion}
While the aforementioned results show promise with regards to implementing real-time fitting algorithms using these techniques, it is worth noting that there is still much opportunity for growth. Algorithm optimization, with regards to kernel weighting, basis function selection, convolution length, and triggering mechanisms based on the filter output will be explored further. Additionally, these techniques can be implemented to perform an \textit{in situ} background subtraction routine to refine more traditional analysis techniques.

Furthermore, while this paper has focused on implementation of a sliding least squares fit, it is worth noting that any general FIR filter can be approximated by piecewise polynomials. Thus arbitrary filter kernels can be developed using traditional methods and then implemented efficiently using the techniques described herein. For example, a simple zero-area cusp with a flat top, which is often considered as an optimal filter for physics applications, is trivially implemented as a three segment filter, of orders two, zero, and two, using the techniques discussed herein.

Lastly, while the focus of this work has been to describe a simple, yet effective, procedure for arbitrary filter implementation via piecewise polynomial recursions on FPGA, the prescribed procedure is not exclusively beneficial for this architecture, nor is it necessarily the most efficient implementation. In the simplest construction of a recursive polynomial convolution kernel of order $K$, the maximum number of multiplications is $2K{+}1$, regardless of the length $L$ of the kernel, compared with $L$ multiplications for a na\"ive sliding scalar product, where $L\sim1000$ is orders of magnitude larger than $K+1\sim 3$. As the algorithm outperforms the efficiency of FFT convolutions of complexity $O(n \log n)$, it may find applications for CPU or GPU as well. Additionally, per Equation \ref{eq:ppsumm}, identical delays for each order of the recursion polynomials are used for ease of mapping between the desired piecewise polynomial fit to the scaled recursion response function. However, as demonstrated in Figure 4 of Jordanov and Knoll \cite{JORDANOV1994337}, choosing unique delay values produces more complex response functions, such that a single recursion can approximate several segments of the desired response function. Using this principle, the authors present a trapezoid filter with 2 accumulators and 1 multiplication, compared to the 5 accumulators and 7 multiplications needed for the generalized approach presented here. However, the difficulty in optimizing these delays to produce the appropriate response function with minimal resource requirements increases with each polynomial order, and the correct optimizations are application specific, so the balance between optimization and time-to-deployment must be considered as needed.

\section{Conclusion}
We have described a method for developing a real-time, non-linear least squares fitting algorithm for digital signal processing in high noise environments. By 1) interpreting rows of the pseudoinverse  of the design matrix as convolution response functions, and then 2) approximating these response functions as  piecewise polynomials, an FPGA-friendly recursive implementation can be developed with high fidelity to the desired template fits, allowing for the exclusion of specific noise patterns through additional template fit functions. Using a synthetic data set, a prototype version of this filter has been compared to commonly used algorithms. The resultant energy and timing characteristics were comparable to those of commonly used timing filters, but a lower triggering threshold was achieved. Furthermore, the filter performance did not suffer as a consequence of low-frequency baseline oscillations, proving its utility as a robust threshold trigger for low-energy proton events.


\ifCLASSOPTIONcaptionsoff
  \newpage
\fi



%
\bibliography{references.bib}

\begin{thebibliography}{10}
\providecommand{\url}[1]{#1}
\csname url@samestyle\endcsname
\providecommand{\newblock}{\relax}
\providecommand{\bibinfo}[2]{#2}
\providecommand{\BIBentrySTDinterwordspacing}{\spaceskip=0pt\relax}
\providecommand{\BIBentryALTinterwordstretchfactor}{4}
\providecommand{\BIBentryALTinterwordspacing}{\spaceskip=\fontdimen2\font plus
\BIBentryALTinterwordstretchfactor\fontdimen3\font minus
  \fontdimen4\font\relax}
\providecommand{\BIBforeignlanguage}[2]{{%
\expandafter\ifx\csname l@#1\endcsname\relax
\typeout{** WARNING: IEEEtran.bst: No hyphenation pattern has been}%
\typeout{** loaded for the language `#1'. Using the pattern for}%
\typeout{** the default language instead.}%
\else
\language=\csname l@#1\endcsname
\fi
#2}}
\providecommand{\BIBdecl}{\relax}
\BIBdecl

\bibitem{8356004}
S.~{Lahti}, P.~{Sjövall}, J.~{Vanne}, and T.~D. {Hämäläinen}, ``Are we
  there yet? a study on the state of high-level synthesis,'' \emph{IEEE
  Transactions on Computer-Aided Design of Integrated Circuits and Systems},
  vol.~38, no.~5, pp. 898--911, 2019.

\bibitem{nakhostin}
M.~Nakhostin, \emph{Signal Processing for Radiation Detectors}.\hskip 1em plus
  0.5em minus 0.4em\relax Hoboken, NJ: Wiley, 2018.

\bibitem{KAMLEITNER201488}
\BIBentryALTinterwordspacing
J.~Kamleitner, S.~Code, S.~Gnesin, and P.~Marmillod, ``Comparative analysis of
  digitial pulse processing methods at high count rates,'' \emph{Nucl. Instrum.
  Methods}, vol. A736, pp. 88--98, 2014. [Online]. Available:
  \url{http://www.sciencedirect.com/science/article/pii/S0168900213013697}
\BIBentrySTDinterwordspacing

\bibitem{JORDANOV1994337}
\BIBentryALTinterwordspacing
V.~Jordanov and G.~Knoll, ``Digital synthesis of pulse shapes in real time for
  high resolution radiation spectroscopy,'' \emph{Nucl. Instrum. Methods}, vol.
  A345, no.~2, pp. 337--345, 1994. [Online]. Available:
  \url{http://www.sciencedirect.com/science/article/pii/0168900294910111}
\BIBentrySTDinterwordspacing

\bibitem{TAN}
H.~Tan, M.~Momayezi, A.~Fallu-Labruyere, Y.~X. Chu, and W.~K. Warburton, ``A
  fast digital filter algorithm for gamma-ray spectroscopy with
  double-exponential decaying scintillators,'' \emph{IEEE Trans. Nucl. Sci.},
  vol.~51, no.~4, pp. 1541--1545, 2004.

\bibitem{ABBIATI}
R.~Abbiati, A.~Geraci, and G.~Ripamonti, ``A new filter concept yielding
  improved resolution and throughput in radiation detection systems,''
  \emph{IEEE Trans. Nucl. Sci}, vol.~52, no.~4, pp. 950--953, 2005.

\bibitem{AGOSTINI}
\BIBentryALTinterwordspacing
M.~Agostini \emph{et~al.}, ``Improvement of the energy resolution via an
  optimized digital signal processing in gerda phase i,'' \emph{Eur. Phys. J.
  C}, vol.~75, p. 255, 2015. [Online]. Available:
  \url{https://link.springer.com/article/10.1140/epjc/s10052-015-3409-6}
\BIBentrySTDinterwordspacing

\bibitem{XIA}
B.~Hubbard-Nelson. (2008) Digital x-ray processor user's manual, revision
  12017. XIA, LLC.

\bibitem{FALLU}
A.~Fallu-Labruyere, H.~Tan, W.~Hennig, and W.~K. Warburton, ``Time resolution
  studies using digital constant fraction discrimination,'' \emph{Nucl.
  Instrum. Methods}, vol. A579, no.~1, pp. 247--251, 2007.

\bibitem{YOUNG}
I.~Young and L.~van Vliet, ``Recursive implementation of the gaussian filter,''
  \emph{Signal Proc.}, vol.~44, no.~2, pp. 139--151, 1995.

\bibitem{TINTORI}
C.~Tintori, ``Digital pulse processing in nuclear physics,'' White paper 2081
  Rev. 2.1, CAEN, 2011.

\bibitem{PETRICK92}
N.~Petrick, A.~Hero, N.~H. Clinthorne, and W.~Rogers, ``Least squares arrival
  time estimators for photons detected using a photomultiplier tube,''
  \emph{IEEE Trans. Nucl. Sci.}, vol.~39, no.~4, 1992.

\bibitem{PETRICK94}
N.~Petrick, A.~Hero, N.~H. Clinthorne, and W.~Rogers, ``A fast least-squares
  arrival time estimator for scintillation pulses,'' \emph{IEEE Trans. Nucl.
  Sci.}, vol.~41, no.~4, 1994.

\bibitem{RIPAMONTI}
G.~Ripamonti and A.~Geraci, ``Towards real-time digital pulse processing based
  on least-mean-squares algorithms,'' \emph{Nucl. Instrum. Methods}, vol. A400,
  p. 447=455, 1997.

\bibitem{JOLY}
B.~Joly, G.~Montarou, J.~Lecoq, G.~Bohner, M.~Crouau, M.~Brossard, and P.~Vert,
  ``An optimal filter based algorithm for pet detectors with digital sampling
  front-end,'' \emph{IEEE Trans. Nucl. Sci.}, vol.~57, no.~1, pp. 63--70, 2010.

\bibitem{474472}
A.~Uritani, O.~Kubota, Y.~Takenaka, and C.~Mori, ``Reduction of microphonic
  noise by digital waveform processing,'' in \emph{Nuclear Science Symposium
  and Medical Imaging Conference, 1994., 1994 IEEE Conference Record}, vol.~2,
  Oct 1994, pp. 905--909 vol.2.

\bibitem{ZIMMERMANN2013404}
\BIBentryALTinterwordspacing
S.~Zimmermann, ``Active microphnoic noise cancellation in radiation
  detectors,'' \emph{Nucl. Instrum. Methods}, vol. A729, pp. 404--409, 2013.
  [Online]. Available:
  \url{http://www.sciencedirect.com/science/article/pii/S0168900213008930}
\BIBentrySTDinterwordspacing

\bibitem{7097420}
V.~Moeller-Chan, T.~Hasenohr, T.~Stezelberger, M.~Turqueti, and S.~Zimmermann,
  ``Microphonic noise cancellation in radiation detectors using real-time
  adaptive modeling,'' in \emph{2014 19th IEEE-NPSS Real Time Conference}, May
  2014, pp. 1--4.

\bibitem{10.5555/1403886}
W.~H. Press, S.~A. Teukolsky, W.~T. Vetterling, and B.~P. Flannery,
  \emph{Numerical Recipes 3rd Edition: The Art of Scientific Computing},
  3rd~ed.\hskip 1em plus 0.5em minus 0.4em\relax USA: Cambridge University
  Press, 2007.

\bibitem{AbramowitzStegun}
M.~Abramowitz and I.~A. Stegun, \emph{Handbook of Mathematical Functions with
  Formulas, Graphs, and Mathematical Tables}.\hskip 1em plus 0.5em minus
  0.4em\relax New York: Dover, 1964.

\bibitem{BUTLER20114076}
\BIBentryALTinterwordspacing
J.~T. Butler, C.~Frenzen, N.~Macaria, and T.~Sasao, ``A fast segmentation
  algorithm for piecewise polynomial numeric function generators,'' \emph{J.
  Comput. Appl. Math.}, vol. 235, no.~14, pp. 4076 -- 4082, 2011. [Online].
  Available:
  \url{http://www.sciencedirect.com/science/article/pii/S037704271100121X}
\BIBentrySTDinterwordspacing

\bibitem{BROUSSARD201783}
\BIBentryALTinterwordspacing
L.~J. Broussard \emph{et~al.}, ``Detection system for neutron $\beta$ decay
  correlations in the {UCNB} and {N}ab experiments,'' \emph{Nucl. Instrum.
  Methods}, vol. A849, pp. 83 -- 93, 2017. [Online]. Available:
  \url{http://www.sciencedirect.com/science/article/pii/S0168900216312943}
\BIBentrySTDinterwordspacing

\bibitem{jezghani}
\BIBentryALTinterwordspacing
A.~P. Jezghani, ``{A DETECTION AND DATA ACQUISITION SYSTEM FOR PRECISION BETA
  DECAY SPECTROSCOPY},'' Ph.D. dissertation, Univ. of Kentucky, Lexington, KY,
  May 2019. [Online]. Available: \url{https://doi.org/10.13023/etd.2019.177}
\BIBentrySTDinterwordspacing

\end{thebibliography}
\bibliographystyle{IEEETran}




%








\end{document}